# Torsional refrigeration by twisted, coiled, and supercoiled fibers

Run Wang[1,#], Shaoli Fang[2,#], Yicheng Xiao[1], Enlai Gao[2,3], Nan Jiang[2,4], Yaowang Li[5], Linlin Mou[1], Yanan Shen[1], Wubin Zhao[1], Sitong Li[1], Alexandre F. Fonseca[6], Douglas S. Galvão[6], Mengmeng Chen[1], Wenqian He[1], Kaiqing Yu[1], Hongbing Lu[7], Xuemin Wang[7,8], Dong Qian[7], Ali E. Aliev[2], Na Li[2,9], Carter S. Haines[2], Zhongsheng Liu[1], Jiuke Mu[2], Zhong Wang[2], Shougen Yin[10], Márcio D. Lima[11], Baigang An[12], Xiang Zhou[13], Zunfeng Liu[1,†], and Ray H. Baughman[2,†]

[1]State Key Laboratory of Medicinal Chemical Biology, College of Pharmacy, Key Laboratory of Functional Polymer Materials, Nankai University, Tianjin, 300071, China.
[2]Alan G. MacDiarmid NanoTech Institute, University of Texas at Dallas, Richardson, TX 75080, USA.
[3]Department of Engineering Mechanics, School of Civil Engineering, Wuhan University, Wuhan, Hubei, 430072, China.
[4]Shenzhen Geim Graphene Center, Tsinghua Shenzhen International Graduate School, Shenzhen 518055, China.
[5]School of Life Sciences, Tsinghua University, Beijing, 100084, China.
[6]Applied Physics Department, State University of Campinas, Campinas, SP 13081-970, Brazil.
[7]Department of Mechanical Engineering, University of Texas at Dallas, Richardson, TX 75080, USA.
[8]Department of Mechanical Engineering, Georgia Southern University, Statesboro, GA, 30458, USA.
[9]Materials Science, MilliporeSigma, Milwaukee, WI 53209, USA
[10]Institute of Materials Physics, Tianjin University of Technology, Tianjin, 300384, China.
[11]Lintec of America, Nano-Science & Technology Center, Richardson, TX 75081, USA.
[12]School of Chemical Engineering, University of Science and Technology Liaoning, Anshan, 114051, China.
[13]Department of Science, China Pharmaceutical University, Nanjing, Jiangsu, 211198, China.

[#]These authors equally contributed to this work.
[†]Corresponding authors. Email: ray.baughman@utdallas.edu (R.H.B) and liuzunfeng@nankai.edu.cn (Z.L.)

**ABSTRACT:** Higher efficiency, lower cost refrigeration is needed for both large and small scale cooling. Refrigerators using entropy changes during cycles of stretching or hydrostatically compression of a solid are possible alternatives to the vapor-compression fridges found in homes. We show that high cooling results from twist changes for twisted, coiled, or supercoiled fibers, including those of natural rubber, NiTi, and polyethylene fishing line. By using opposite chiralities of twist and coiling, supercoiled natural rubber fibers and coiled fishing line fibers result that cool when stretched. A demonstrated twist-based device for cooling flowing water provides a high cooling energy and device efficiency. Theory describes the axial and spring index dependencies of twist-enhanced cooling and its origin in a phase transformation for polyethylene fibers.

**Summary:** Twist-exploiting mechanocaloric cooling is demonstrated for rubber fibers, fishing line fibers, and NiTi shape-memory wires.

Compared with conventional vapor compression refrigerators, a solid that changes entropy when deformed could possibly provide a higher efficiency; lower cost, weight, and volume; and more convenient miniaturization. An ordinary rubber band is an example, which was reported in 1805 to heat when stretched and to cool when stretch is released (*1*). Alternative materials for refrigeration are electrocaloric (*2–6*) and magnetocaloric (*5, 7*), meaning they provide cooling because of changes in electric or magnetic fields, respectively. While these solid-state cooling mechanisms have been widely investigated for diverse materials, none yet meets the performance needs for wide scale applications. In fact, thermoelectrics are the only solid-state coolers that have been commercialized, and these coolers are expensive and have low energy conversion efficiencies.

Efficiencies of about 60% of the Carnot efficiency can be obtained for vapor compression refrigerators (*8, 9*), which is a mature technology that typically uses gases whose release can powerfully contribute to global warming. No alternative technology presently achieves this efficiency. On the other hand, theoretical efficiencies of about 84% of the Carnot efficiency have been predicted for tensile deformation cycles of a NiTi shape memory alloy (*9*). Realization of close to this efficiency could have enormous consequences if the needed cycle life of vapor compression refrigerators could be obtained, since so much of global energy consumption goes to cooling.

Refrigeration materials that undergo cooling because of any type of mechanical deformation are called mechanocaloric, and those that cool because of changes in uniaxial stress (*5–15*) or hydrostatic pressure (*5, 15–17*) are more specifically called elastocaloric and barocaloric, respectively (*6*). We here demonstrate cooling by a change in yarn or fiber twist, which we call twistocaloric cooling. We call coolers that use twist changes for refrigeration "twist fridges".

**Twist-based cooling using single natural rubber fibers**

Vulcanized natural rubber (NR) fibers (*18*, figs. S1−S10) provide a prototypical twistocaloric material. Twisting these fibers causes coil nucleation and propagation, decreasing inter-coil separation, and finally nucleation of supercoils (Fig. 1A). For isometric strains (i.e., constant strains during twist) of 100%, 250%, 300%, and 450%, fracture occurs before completion of supercoiling, initiation of supercoiling, completion of coiling, and initiation of coiling, respectively. Unless otherwise stated, (a) stresses, strains, and twist densities are relative to the cross-sectional area or length of the non-deformed fiber (called the parent fiber) and (b) a tensile strain rate of 42 cm/s and a twisting/untwist rate of 50 turns/s were used for NR fibers having a non-deformed length of 3.0 cm. Here and elsewhere, tensile strain changes are made either while a fiber or yarn is torsionally tethered or for the non-twisted state.

Twistocaloric surface temperature changes occur inhomogeneously for coiled and supercoiled fibers. Hence, maximum ($\Delta T_{max}$) and average ($\Delta T_{avg}$) surface temperature changes are reported. Surface temperatures were measured using a thermal camera, whose accuracy was established by comparison with thermocouple measurements in a temperature-controlled oven on fibers having different twist states (*18*).

The temperature swing on stretch and stretch release (between heating and cooling temperatures) is important for remotely optically readable strain sensors or mechanothermochromic fibers. For 300% isometric strain, this swing in surface-average temperature was 3.3, 6.5, and 7.7°C for highly twisted, coiled, and partially supercoiled fibers, respectively, compared to 2.4°C for a non-twisted fiber (Fig. 1B). The maximum surface temperature swing (fig. S11B) for this modest applied strain was 5.4, 10.5, and 12.9°C for the highly twisted, coiled, and partially supercoiled fibers, which is up to 5.4 times that for the non-twisted NR fiber. However, when released from strains above 300%, non-twisted NR fibers provided greater cooling. Hence, these twisted NR fibers are not very useful for refrigerators that solely use large-magnitude stretch and release.

Examples of cooling by untwist are shown in Fig. 1, C−E, figs. S13−S18, and Movie S1. When twist is released from a supercoiled NR fiber having 100% isometric strain, the maximum (−15.5°C) and average (−12.4°C) surface cooling (Fig. 1B) exceed that for 600% stretch release from a non-twisted fiber (−12.2°C). By releasing twist and then stretch (Fig. 2, A and B), even higher maximum (−16.4°C) and average (−14.5°C) cooling was obtained. Hence, a twist-based cooler at 100% strain can be 2/7$^{th}$ the maximum length of the above stretch-based cooler, and still provide 2.3°C higher surface-average cooling. Figure S34A and Movie S2 demonstrate the use of a thermochromic-dye-coated NR fiber as a mechanothermochromic material whose color responds to changes in inserted twist.

The heating and cooling during stretch and stretch release of a non-twisted fiber are approximately independent of fiber diameter (fig. S11A). Also, NR fibers having different diameters and the same elongation have approximately the same dependence of $\Delta T_{max}$ and $\Delta T_{avg}$ on the product of twist density and stretched fiber diameter (fig. S17). The maximum surface cooling on twist release was essentially the same whether twist insertion was done isometrically (for constant tensile strain) or isobarically (for constant tensile load), as long as the percent stretch obtained after isometric twist release was identical (fig. S19). The NR fiber showed stable heating/cooling during isometric twisting/untwisting up to 15 turns/cm for 750 cycles (fig. S21).

**Volume-average cooling for twisted NR fibers**

The volume-average cooling upon untwisting a NR fiber was obtained from the cooling energy of this fiber in a water-containing plastic tube (fig. S44A). This cooling energy is the sum of the products of gravimetric heat capacity, mass, and temperature decrease for the water, plastic tube, and NR fiber after internal thermal equilibration of the system (fig. S44B). The volume-average cooling upon untwisting a NR fiber was obtained by dividing the derived cooling energy by the heat capacity and mass of the NR fiber (*18*). To minimize loss of cooling energy to surrounding air (fig. S12) by decreasing the temperature change of the water, the mass of water was 4.63-times that of the NR fiber. This small loss of cooling energy was corrected by fitting the long-term temperature dependence of water and container to a standard heat-loss equation (*19*), and then adding these temperature corrections to the observed curves, so that the loss-corrected temperature curves reach long plateaus (*18*). The temperature changes for these plateaus were used to obtain the cooling energy, and thereby the dependence of volume-average temperature change on twist density change.

Figure 1D compares the surface-average and volume-average temperature changes for isometric twist insertion and release from a 100%-stretched NR fiber that is partially coiled or supercoiled. For the high twist density most important for applications, where the twist removal time (1.5 s) exceeds the calculated radial thermal equilibration time (0.96 s, *18*), the volume-average cooling is 11.3°C and the ratio of volume-averaged to surface-averaged temperatures reaches 0.91. This is consistent with nearly complete equilibration in the radial direction.

Although the maximum twistocaloric specific cooling energy obtained (19.4 J/g) is lower than reported for releasing 600%-stretch from a non-twisted NR fiber (21.6 J/g) (*20*), the stretch needed for this torsional cooling is much smaller (100%). At 100% strain, stretch-based cooling would only be 4.2% of that delivered by twistocaloric cooling (*18*).

**Coefficients of performance for twisted NR fibers**

Using the above volume-average cooling on isometric twist release, the coefficient of performance (COP) for NR twist fridges was obtained. The COP can be described as the ratio of the cooling energy (the product of the volume-average cooling, the gravimetric heat capacity, and the mass of the fiber) to either the input mechanical energy or the net energy consumed during a mechanical cycle (called $COP_{HC}$ and $COP_{FC}$, for a half-cycle and a full-cycle, respectively) (*5*). $COP_{HC}$ is useful when device simplicity or miniaturization is more important than increasing efficiency by recapturing part of the input mechanical energy.

Using the stretch-release-induced cooling of the non-twisted fiber (Fig. 1B), the volume-average cooling during twist release (Fig. 1D), and mechanical measurements (figs. S7 and S10B), the COPs were obtained as a function of volume-average cooling (Fig. 2C and fig. S43). The $COP_{FC}$ and $COP_{HC}$ for cooling, either by twist release or by releasing twist and then stretch, are much higher than for cooling by stretch release from the non-twisted fiber when the volume-average cooling is above −0.4°C (Fig. 2C and fig. S43). For the highest volume-average cooling obtained for releasing stretch from a non-twisted fiber (−12.2°C), isometric untwist (−11.3°C), and releasing twist and then stretch (−12.1°C), the $COP_{FC}$ values are 3.8, 8.3, and 8.5, respectively (Fig. 2C), and the $COP_{HC}$ values are 1.9, 5.2, and 5.3, respectively (fig. S43).

The intrinsic efficiency of a material, which is the maximum theoretical efficiency for a refrigerator, is the ratio of $COP_{FC}$ to the COP of a Carnot cycle ($COP_{Carnot}$). $COP_{Carnot}$ is $T_C/(T_H-T_C)$, where $T_C$ and $T_H$ are the minimum and the maximum temperatures in the cycle, respectively. These material efficiencies for cooling during isometric untwisting and during releasing twist and then stretch are much higher than for cooling during stretch release from a non-twisted fiber, for both full-cycle and half-cycle processes (fig. S43, C and D). For the highest volume-average cooling for stretch release from a non-twisted fiber, isometric untwisting, and releasing twist and then stretch, the full-cycle intrinsic material efficiencies were 0.32, 0.63, and 0.67, respectively.

**Twist-based cooling by plying natural rubber fibers**

To demonstrate scalability, cooling was measured while isometrically unplying multiple NR fibers. Unplying seven-ply 2.2-mm-diameter NR fibers (Fig. 2, D and E) produced higher maximum (–19.1°C) and average (–14.4°C) surface cooling than for 600% stretch release from a non-twisted fiber (–12.2°C, Fig. 1B) or for isometric untwisting of a 100%-stretched single fiber (–15.5°C maximum and –12.4°C average, from Fig. 1C and fig. S15). Unplying from isometric strains around 100% produced the highest twistocaloric cooling (figs. S23 and S24).

Guided by the correspondence between twistocaloric cooling for single fibers having the same stretch, and a similar product of twist density and stretched fiber diameter (fig. S17), we found that the twistocaloric cooling associated with isometric fiber plying approximately depends on the product of the twist density of plying and the effective diameter of the bundle ($D_{eff} = n^{0.5} \times D_s$, where $n$ is the number of fibers and $D_s$ is the stretched fiber diameter) (Fig. 2E).

**Spatial and temporal dependencies of twistocaloric temperature changes for NR fibers**

Spatially complex surface temperature changes occur for twistocaloric processes for coiled or super-coiled fibers (Fig. 2F and fig. S20). Like for a bent cantilever, the inside of a coiled fiber is compressed and the outside is stretched, causing the coil's exterior to undergo the highest temperature change during coil formation and removal. Painting the exterior of a coiled fiber enabled demonstration that these maximum strained regions have the highest cooling after fiber untwisting. The periodicity of these regions on the surface of the untwisted fiber (5.6 mm) is much longer than the average coil period (2.5 mm) of the fully coiled fiber.

The peak surface cooling increases with increasing distance from the site where coiling nucleates (fig. S20), since coiling stretches non-coiled fiber regions and thereby decreases the spring index (the ratio of the average coil diameter to the diameter of the fiber within the coil) of later-introduced coils. The sharp change in maximum surface cooling upon complete removal of coiling (fig. S17A) disappears when measuring average cooling (fig. S17B), indicating that the effects of bending approximately average to zero. Since the number of coils is strain invariant, the separations between temperature peaks during both stretching and releasing are identical to the inter-coil separations (Fig. 1B inset).

**Twistocaloric cooling by self-coiled polyethylene and nylon fibers**

Twistocaloric cooling was also investigated for polymers used for fishing line and sewing thread. These polymers were made elastically deformable by isobarically twisting until the fiber coils, as is done to make powerful thermally-driven artificial muscles (*21*). These self-coiled fibers are called homochiral, because fiber twist and coiling have the same handedness.

Initial experiments employed 0.41-mm-diameter, 65-pound test, braided polyethylene (PE) fishing line (fig. S25A). Spring indices from 1.4 to 0.5 were obtained by inserting 6.5 to 7.3 turns/cm of twist under isobaric loads from 37.1 to 74.2 MPa, respectively. When releasing 22.7% strain, maximum and average surface cooling of –5.1°C and –3.2°C, respectively, were obtained for a spring index of 0.8 (Fig. 3A and fig. S25). For comparison, the temperature change upon stretch release of non-twisted polyethylene yarn was below ±0.1°C, and the highest reported elastocaloric cooling for a non-elastomeric polymer is –2.5°C for a poly(vinylidene fluoride-trifluoroethylene-chlorotrifluoroethylene) terpolymer (*15*), which is a relaxor ferroelectric (*2*).

The highest temperature swing between maximum heating and cooling (12.9°C, for a spring index of 0.8) was obtained for 22.7% stretch and stretch release, which is much higher per strain change (0.57°C/%) than for a non-twisted NR fiber (0.04°C/% for 600% strain). This high sensitivity is useful for mechanothermochromic indicators (Movie S3). No degradation in performance occurred during the investigated 2500 stretch/release cycles to 13% strain (fig. S26). The twistocaloric cooling for self-coiled, high-strength polyethylene yarn at 10°C ambient temperature is smaller than for ambient temperatures of 25 and 40°C (fig. S25). In contrast, the stretch-release-induced cooling for a non-twisted NR fiber at 10°C is higher than for either higher or lower ambient temperatures (*22*).

Twistocaloric cooling was also observed during stretch release for coiled, single-filament nylon 6 fishing line having parent diameters (*D*) of 0.2, 0.4 and 0.6 mm. So that the fibers have the same spring index (1.0), they were coiled using the same isobaric stress and the same twist number (twist density times *D*). These homochiral fibers, having progressively larger diameters, provided progressively increasing maximum cooling (–1.3, –1.9, and –2.1°C) and average cooling (–0.8, –1.2, and –1.8°C) upon stretch release (fig. S28).

The strain dependence of twistocaloric cooling generally increases with decreasing spring index for coiled, high-strength polyethylene yarn (fig. S25), low-strength polyethylene (fig. S27), and nylon 6 fibers (fig. S28). Similarly, for coiled polymer muscles (*21*), higher heating is needed to cause a given stroke for a smaller spring-index muscle. The origin in both cases is the stretch-induced twist change per fiber length, whose magnitude is spring-index dependent.

Theoretical results (*18*, fig. S39) show that the stretch-induced twist change per fiber length, divided by the percent stretch of a coiled fiber, should approximately depend on the inverse square of coil spring index. This dependence arises since both the strain needed to pull out one coil and the fiber length per coil linearly increase with coil diameter. In agreement, the observed twistocaloric temperature changes for self-coiled polyethylene and nylon 6 fibers or yarns are approximately proportional to the ratio of percent stretch to the square of the spring index (Fig. 3A and fig. S40). This dependency is predicted only for the case where different-spring-index yarns have approximately the same initial coil bias angle. Due to the difficulty of controlling this coil bias angle for later-described mandrel-coiled fibers, deviations from these theoretical results are observed.

**Origin of twistocaloric cooling of polyethylene fibers**

Since the above polyethylene and nylon 6 fibers are crystalline, it is challenging to explain the origin of the entropy decrease caused by stretching the coiled fiber. One possible explanation for polyethylene is the known deformation-driven orthorhombic-to-monoclinic phase conversion (fig. S41) (*23−26*). Using molecular dynamics calculations at 300 K (*18*), we predict that the entropy of the orthorhombic phase is 0.097 $JK^{-1}g^{-1}$ higher than for the monoclinic phase, which agrees with the entropy difference calculated over thirty years ago (*26*) from vibrational spectra (~0.12 $JK^{-1}g^{-1}$). Our predicted free energy difference due to this entropy change (29.1 J/g), and the specific heat capacity (*27*) of polyethylene (1.56 $JK^{-1}g^{-1}$) provides a predicted −18.7°C cooling if coil stretch caused complete conversion from the orthorhombic to monoclinic phase.

X-ray diffraction (XRD) measurements for polyethylene show that self-coiling results in partial conversion of orthorhombic to monoclinic phase. Thermally annealing the coiled yarn (120°C for 2h) largely reverses this transformation (Fig. 3B and fig. S42B), but little affects twistocaloric cooling (fig. S42A). Stretching the coiled yarn to 20% strain reversibly increases the amount of monoclinic phase from 5.0% to 11.4% (Table S2). Using this percent change of monoclinic phase and the −18.7°C cooling for 100% orthorhombic-to-monoclinic phase conversion, a volume-average cooling of −1.2°C is predicted for 20% strain release, which is identical to the measured surface-average cooling for this strain release. This suggests that monoclinic-to-orthorhombic phase conversion importantly contributes to twistocaloric cooling. The entropy change contribution from the amorphous phase is apparently small, which is consistent with the negligible change in amorphous scattering profile and with the unchanged ratio of the integrated diffraction intensity of the amorphous phase to the total diffraction intensity for the crystalline phases (from 0.49 to 0.50 during 20% stretch).

**Twistocaloric cooling by stretching heterochiral fibers**

While stretch increases fiber twist for a homochiral coil, stretch causes fiber untwist for a heterochiral coil, which enables reversal of the temperature changes produced by stretch and stretch release. A heterochiral NR coil was made by wrapping a self-coiled 2.2-mm-diameter NR fiber onto a mandrel to provide a spring index of 2.5 (*18*). This mandrel was retained during twistocaloric measurements to prevent twist cancelation in the rubber fiber.

This heterochiral NR coil provides an inverse mechanocaloric effect. For 200% strain, the maximum surface cooling during stretch was –0.8°C, and the maximum surface heating upon stretch release was +0.5°C (Fig. 1F and fig. S22A). For comparison, stretching and releasing the same strain from an identical spring index homochiral coil caused maximum surface temperature changes of +0.5°C and –0.5°C, respectively (Fig. 1F and fig. S22B).

An inverse twistocaloric effect was also found for heterochiral coils of nylon 6 and polyethylene, which can be thermally set to maintain their coiled shape without retention of a mandrel. Stretching heterochiral coils of nylon 6 and polyethylene, with spring indices of 2.0, induces cooling of –0.6 and –0.3°C, respectively, and stretch release produces heating of +0.5 and +0.5°C, respectively (Fig. 3C, figs. S31 and S32). For comparison, stretching identical spring index homochiral coils of nylon 6 and polyethylene produces a maximum heating of +0.8 and +1.4°C, respectively, and stretch release produces a maximum cooling of −0.5 and −0.6°C, respectively (Fig. 3C, figs. S29 and S30). When a heterochiral nylon 6 fiber was stretched to 90%, which is beyond the coil's yield stress, the coil irreversibly deformed to form segments having short (1.0 mm) and long (2.8 mm) inter-coil periods, which cooled and heated during stretch, respectively (fig. S33).

**Twistocaloric cooling by twisting and plying NiTi wires**

Large, reversible temperature changes also result from twist insertion and removal from a single NiTi shape memory wire, and from plying and unplying these wires. The investigated 0.7-mm-diameter $Ni_{52.6}Ti_{47.4}$ wires had a martensite-to-austenite transition from −29.0 to 15.0°C during heating, and an austenite-to-martensite transition from 13.5 to −44.6°C during cooling (fig. S35B).

Twistocaloric cooling resulted from isometrically untwisting a single NiTi wire (–17.0°C maximum and –14.4°C average surface temperature changes) at 0% strain and at a twist rate of 50 turns/s (Fig. 4A and figs. S36 and S37). Unplying a four-ply bundle of NiTi wires produced even higher cooling (–20.8°C maximum and –18.2°C average surface temperature changes), compared with the –17.0°C elastocaloric cooling provided by an identical NiTi wire (Table S1). This twistocaloric cooling was stable over 1000 cycles of twisting/untwisting up to 0.6 turns/cm (fig. S38). XRD results demonstrate reversible austenite/martensite conversion during twisting/untwisting a NiTi wire (Fig. 4B). Since this conversion is incomplete, there is an opportunity to increase twist-induced cooling by optimizing the NiTi composition and the operating temperature range of the cooler.

The calculated time for internal thermal equilibration of a 0.7-mm-diameter NiTi wire was 2.5 ms (*18*), which is much faster than the untwist process at 15 turns/s (65 to 455 ms). In agreement, the optically-measured surface-average cooling upon untwisting a NiTi wire is close to the volume-average cooling (Fig. 4C) derived from the calorimetric specific cooling energy of this wire (*18*, fig. S45). The anomalous behavior at very high twist, where cooling slightly decreases, is likely due to the onset of wire buckling. The maximum specific cooling energy (7.9 J/g) for untwisting the NiTi wire (Fig. 4C) is similar to literature results (5.0 to 9.4 J/g) for stretch-released NiTi wires and sheets and for stretch release of the present wires (9.4 J/g) (*18*).

A device that enabled one cycle of refrigeration was demonstrated for the cooling of a stream of water (Fig. 4D and figs. S46 and S47). Flowing ambient-temperature water over a three-ply NiTi wire cable, while removing 0.87 turns/cm of plying, cooled the water by up to

–4.7°C. Higher water cooling (-7.7 °C) resulted from adding thermal insulation, increasing the water channel diameter, and increasing the water flow rate (*18*). Integrating the cooling until the water stream returned to ambient temperature provided a specific cooling energy of 6.75 J/g (Table S3), indicating that much of the cooling energy of the NiTi wire has effectively chilled the water. The twist-produced temperature changes of a thermochromic-paint-coated NiTi wire provided visible indication of torsional rotation (fig. S34B, Movie S4).

**Summary**

Twist release from fibers or fiber plies resulted in high cooling for materials as different as natural rubber and NiTi. The material cooling efficiency was doubled (to 65%) and cooler length was reduced by a factor of 2/7 by replacing elastocaloric cooling by twistocaloric cooling for NR fibers. A NiTi-based twist-fridge cooled flowing water by up to −7.7°C in one cycle.

Twist induced by stretch release from coiled polyethylene fishing line resulted in over 50 times higher surface cooling than obtained by releasing stretching from the non-twisted fiber. Depending upon the relative chiralities of fiber and coil, polymers were engineered to cool either during stretch or stretch release. The spatial periodicity of surface temperature changes of coiled fibers can be an asset for remotely readable tensile and torsional strain sensors, and for color-changing fibers for fabrics that dynamically respond to body movement.

**ACKNOWLEDGEMENTS**

**Funding:** Support from China was from the National Key Research and Development Program of China (grant 2017YFB0307000), the National Natural Science Foundation of China (grants U1533122, and 51773094), the Natural Science Foundation of Tianjin (grant 18JCZDJC36800), the Science Foundation for Distinguished Young Scholars of Tianjin (grant 18JCJQJC46600), the Fundamental Research Funds for the Central Universities (grant 63171219), and the State Key Laboratory for Modification of Chemical Fibers and Polymer Materials, Donghua University (grant LK1704). AFF and DSG are fellows of the Brazilian Agency CNPq (grants 311587/2018-6 and 307331/2014-8, respectively) and acknowledge support from São Paulo Research Foundation (FAPESP) (grants 2018/02992-4 and FAPESP/CEPID 2013/08293-7, respectively). Support at the University of Texas at Dallas was from Air Force Office of Scientific Research grant FA9550-18-1-0510, the Robert A. Welch Foundation grant AT-0029, the National Science Foundation (grants CMMI-1661246, CMMI-1636306, CMMI-1726435, and CMMI-1727960), and the Louis Beecherl Jr. endowed chair. This paper is dedicated in celebration of the 100[th] anniversary of Nankai University. **Author contributions:** R.W., S.F.,

**Figures and Figure Legends**

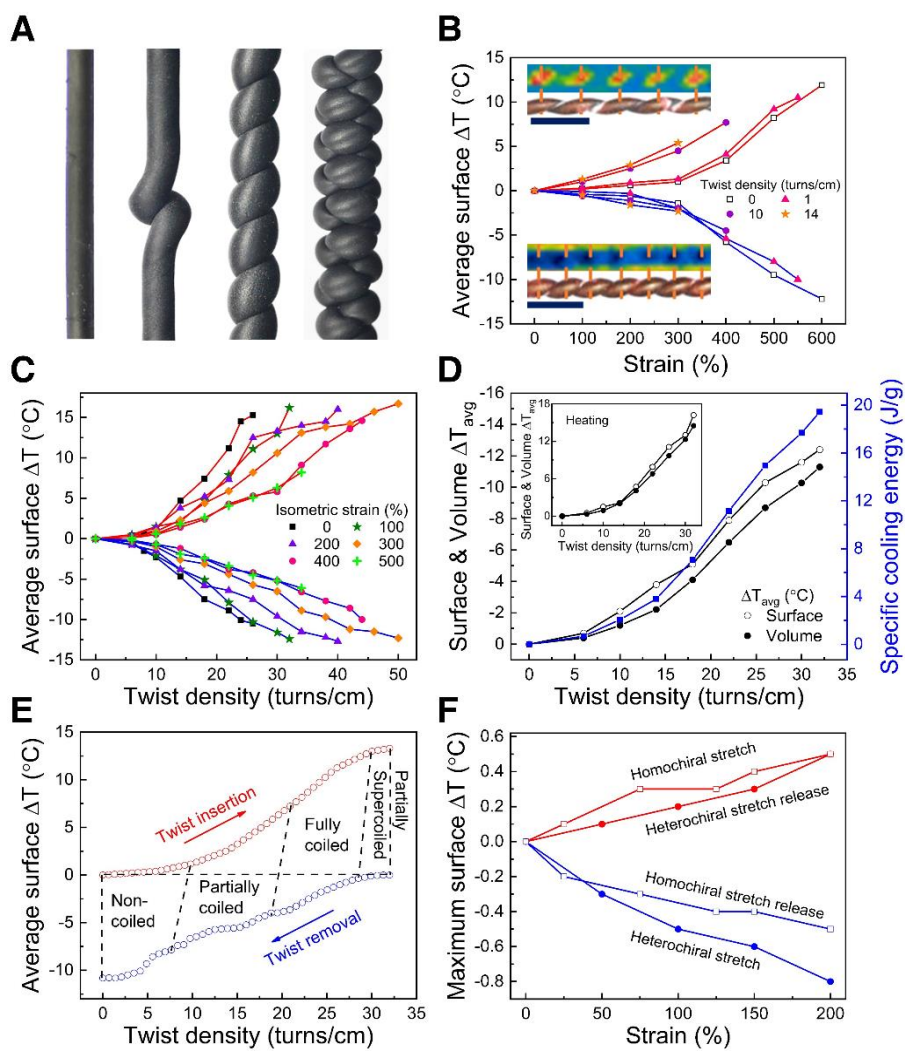

**Fig. 1. Twist-based mechanocaloric performance for single rubber fibers.** (**A**) Photographs of twisted, partially coiled, fully coiled, and fully supercoiled 2.5-mm-diameter NR fibers at 100% strain. The background was digitally removed. (**B**) Surface-average temperature changes versus strain for NR fibers having 0, 1, 10, and 14 turns/cm of twist, which were non-twisted, highly-twisted, fully coiled, and partially supercoiled, respectively, in the non-stretched state. Insets: optical and thermal photographs after 100% stretch (top) and after stretch release (bottom) for a coiled NR fiber (18 turns/cm). The scale bars are 0.5 mm. (**C**) Surface-average temperature changes of a NR fiber during isometric twist and untwist at different tensile strains. (**D**) Surface-average cooling, volume-average cooling, and specific cooling energy for isometric untwisting a NR fiber at 100% strain. Inset: Corresponding results for volume-average and surface-average heating. (**E**) Surface-average temperature changes for isometrically twisting and untwisting a NR fiber at 200% strain, when measured continuously during an increasing twist and a decreasing twist scan. (**F**) Twistocaloric temperature changes versus strain for heterochiral and homochiral coiled NR fibers. The experiments of (B) to (F) used 2.2-mm-diameter NR fibers.

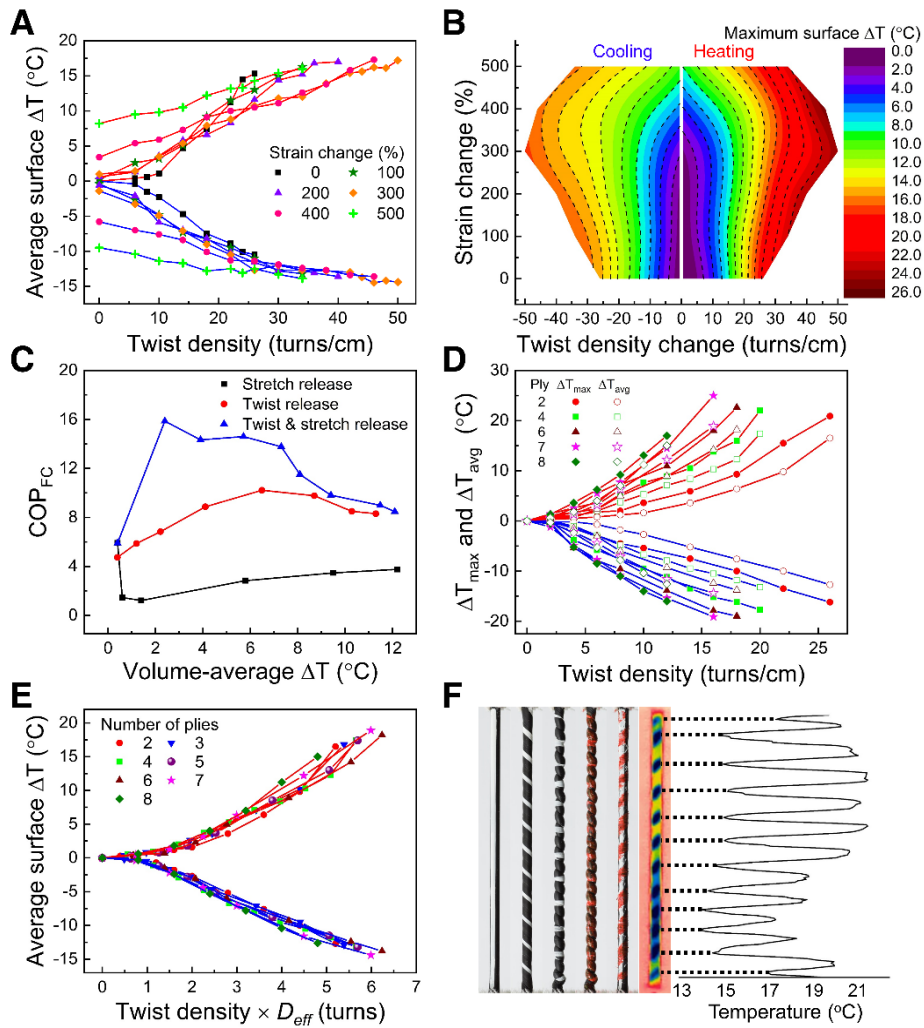

**Fig. 2. Mechanocaloric performance of single and plied NR fibers during twist and stretch.** (**A**) The average and (**B**) maximum surface temperature changes of a NR fiber during sequential stretch and then isometric twisting and during sequential isometric untwisting and then stretch release. (**C**) The COP$_{FC}$ versus volume-average cooling for releasing up to 600% strain from a non-twisted fiber (black), isometric untwisting a fiber at 100% strain (red), and isometric untwisting at 100% strain followed by stretch release (blue). (**D**) The maximum and surface-average temperature changes versus twist density for plying/unplying NR fibers at 100% strain. (**E**) The surface-average temperature changes of (D) versus the product of twist density and the effective fiber diameter. (A) to (E) used 2.2-mm-diameter NR fibers. (**F**) From left-to-right, photographs of an initially non-twisted, 3-cm-long, 3-mm-diameter NR fiber that was: stretched to 100% and painted with a white line along its length; highly twisted; fully coiled; painted red on the coil's exterior; and then fully untwisted. The thermal image and temperature profile (right) show the fiber immediately after removing 12 turns/cm of inserted twist.

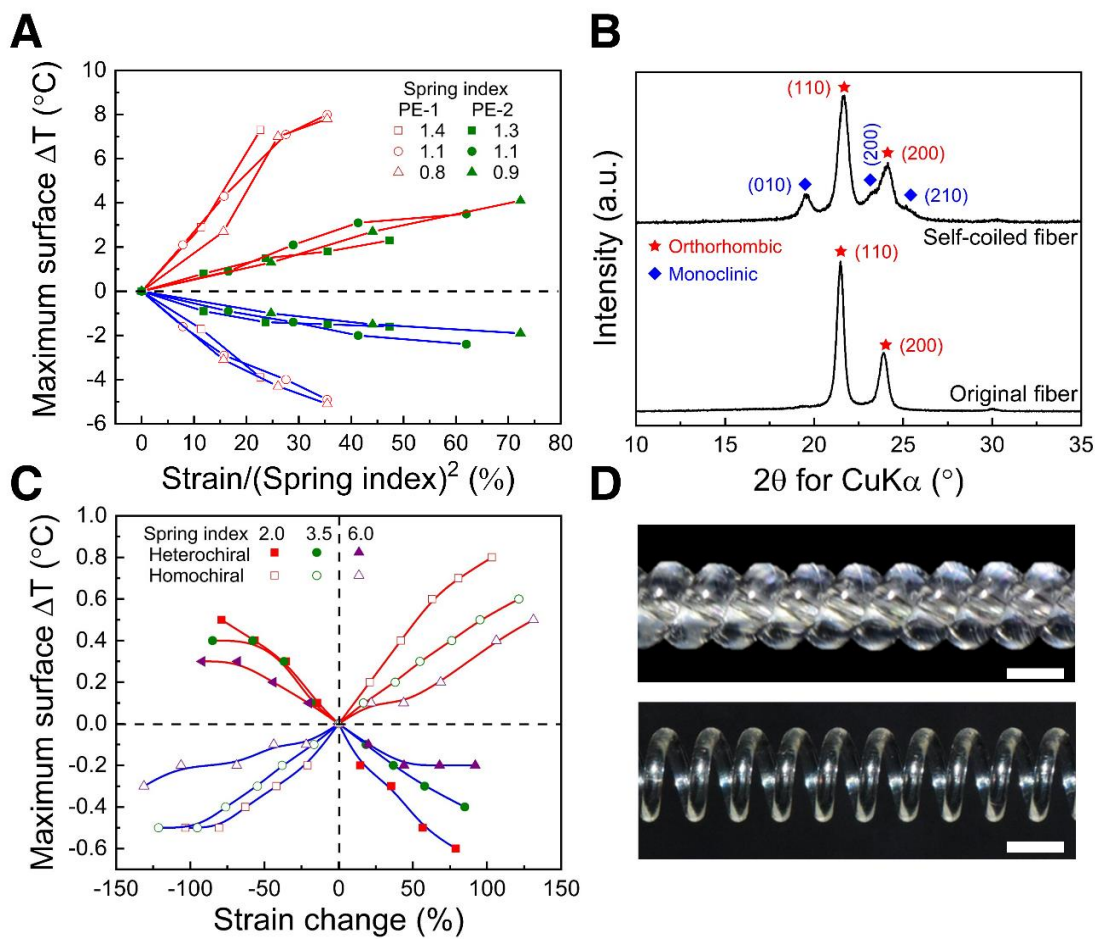

**Fig. 3. Twistocaloric performance for coiled polyethylene and nylon 6.** (**A**) The maximum changes in surface temperature versus the ratio of tensile strain to the square of spring index, for high-strength (PE-1) and low-strength (PE-2) polyethylene fibers. (**B**) XRD for non-twisted and self-coiled PE-1 fibers. (**C**) Twistocaloric temperature changes versus strain for homochiral and heterochiral coiled nylon 6 fibers having different spring indices. (**D**) Optical images of a self-coiled, 0.6-mm-diameter nylon 6 fiber (top) and a thermally annealed (180°C for 1h), mandrel-coiled, 0.6-mm-diameter nylon 6 fiber. Scale bars: 1.0 mm (top) and 2.0 mm (bottom).

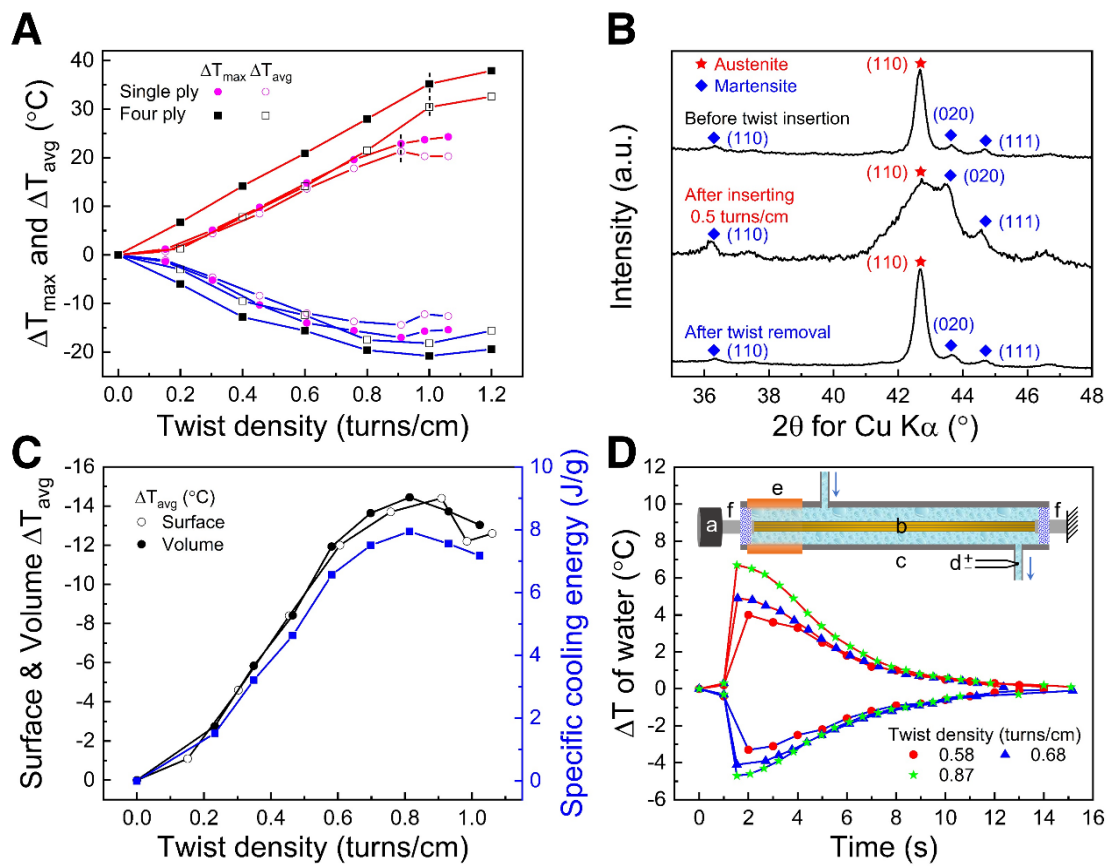

**Fig. 4. Twistocaloric performance for twisted and plied NiTi wires.** (**A**) The twist-induced maximum and surface-average temperature changes for a single NiTi wire and four-ply wires at 0% nominal strain. The dashed lines indicate where buckling occurs. (**B**) XRD for a NiTi wire before twist insertion, after twist insertion, and after twist removal. (**C**) Surface-average cooling, volume-average cooling, and specific cooling energy for untwisting a 0.7-mm-diameter NiTi wire at 0% strain. (**D**) The time dependence of outlet water temperature change following isometric twist insertion and twist removal for a three-ply, 0.6-mm-diameter NiTi wire at a water flow rate of 0.04 mL/min using the device in the inset. This device contains (a) a motor, (b) NiTi wires, (c) a PP tube containing flowing water, (d) a thermocouple, (e) a rubber tube, and (f) epoxy resin that seals tube ends.